\documentclass[letterpaper,twocolumn,10pt]{article}

\usepackage{usenix}
\usepackage{epsfig}
\usepackage{endnotes}
\usepackage[utf8]{inputenc}
\usepackage{url}
\usepackage{hyperref}
\usepackage{listings}
\usepackage{amssymb}
\usepackage{amsfonts}
\usepackage{amsmath}
\usepackage{tabularx,calc}
\usepackage{rotating}
\usepackage{graphicx}
\graphicspath{{figures/}}

\usepackage{subfigure}
\usepackage{algorithm}
\usepackage{algorithmic}
\usepackage{verbatim}
\usepackage{booktabs}
\usepackage{multirow}

\usepackage{comment}
\usepackage[usenames,dvipsnames]{color}
\usepackage{xspace}
\usepackage[normalem]{ulem}
\usepackage{array}
\usepackage{wrapfig}
\usepackage{syntax}
\usepackage{authblk}
\usepackage{flushend}

\usepackage{titlesec}
 \titlespacing{\section}{0pt}{*1.0}{*1.0}
 \titlespacing{\subsection}{0pt}{*0.7}{*0.7}
 \titlespacing{\subsubsection}{0pt}{*0.6}{*0.6}

\renewcommand{\paragraph}[1]{\smallskip\noindent\textbf{#1}}

\newcommand{\todo}[1]{\textbf{[TODO: #1]}}
\newcommand{\ourcomment}[2]{\textbf{[#1: #2]}}
\newcommand{\takeout}[1]{}
\newcommand{\appname}{\textsc{Mavis}\xspace}

\newcommand{\eg}{{\em e.g.}}
\newcommand{\ie}{{\em i.e.}}

\setboolean{ALC@noend}{true}

\algsetup{indent=.7em} 

\InputIfFileExists{locals}{}{}
\begin{document}

\title{\Large \bf \appname: Managing Datacenters using Smartphones}

\newcommand{\authorind}{~~~~~~~~~~~~~~~~~~~~~~~~~~~~~}
\newcommand{\authornegind}{~~~~~~~~~~~~}
\newcommand{\authornegindtwo}{~~~~}

\author[$\dagger$]{Raghav Shankar}
\author[$\dagger$]{Benjamin Kobin}
\author[$\dagger$]{Saurabh Bagchi}
\author[$\ddagger$]{Michael Kistler}
\author[$\ddagger$]{Jan Rellermeyer}
\affil[$\dagger$]{Purdue University}
\affil[$\ddagger$]{IBM Research}

\renewcommand\Authands{ and }

\maketitle
\pagestyle{plain}

\subsection*{Abstract}

Distributed monitoring plays a crucial role in managing the activities of cloud-based datacenters. System administrators have long relied on monitoring systems such as Nagios and Ganglia to obtain status alerts on their desktop-class machines. However, the popularity of mobile devices is pushing the community to develop datacenter monitoring solutions for smartphone-class devices. Here we lay out desirable characteristics of such smartphone-based monitoring and identify quantitatively the shortcomings from directly applying existing solutions to this domain. Then we introduce a possible design that addresses some of these shortcomings and provide results from an early prototype, called \appname, using one month of monitoring data from approximately 3,000 machines hosted by Purdue's central IT organization. 

\section{Introduction}
Maintaining and managing cloud datacenters is a crucial task for cloud vendors because if a cloud vendor does not provide a high degree of availability for the machines in its datacenters, the applications hosted on them will have outages, and this could lead to loss of reputation and revenue for the cloud vendor. 
In order to maintain and manage datacenters, administrators monitor a variety of performance metrics from various levels of the software stack, such as utilization of CPU, memory, and network resources, to (higher in the stack) the frequency of garbage collection in a Java or C\# runtime. There exist tools to aggregate raw data and create higher-level information so that the admins can take prompt action. Nagios \cite{nagios} and Ganglia \cite{ganglia} are two of the most popular commercial systems which are utilized today for systems management. They can provide raw metric values and also compare values against thresholds to generate alerts for simple subscriptions. They also provide graphical outputs for human consumption. 

While systems management through desktop interfaces is the norm today, increasingly the need is being felt for performing systems management through mobile devices \cite{meraki-systemsmanager, bagchi2013lilliput, aNag}. This need is arising increasingly often due to three reasons --- the first is the ubiquity of smart phones, the second (caused in part by the first) is the mobile workforce of today, and the third is the need for responding promptly to outages or impending outages in the cloud infrastructure. The second factor speaks to the fact that the workforce is no longer tethered to desktop-class machines for all of their workdays. 

Let us first consider two central requirements in monitoring datacenter activities. 
First, the monitoring data must be processed quickly and anomalous or suspect events must be detected quickly.
Second, the overhead of monitoring must be kept to a bare minimum on the machines in the datacenter, so that they can perform their main function as the workhorses for running the workloads.
 In addition to these, performing systems management through mobile devices has some specialized, and equally crucial, challenges. First, the mobile devices are resource constrained, significantly in battery and bandwidth. They cannot be relied on to gather huge amounts of data and perform sophisticated data analysis on the device. 
Second, the tools available on mobile devices are quite limited and some of the available ones do not match their user interface to the constraints of mobile devices. For example, some create extensive graphs that may be difficult to discern and interpret on mobile devices. 

\noindent {\bf Our position}: {\em We argue that a three-pronged approach addresses the above-mentioned problems of systems management from mobile devices.} {\em First}, an intermediate server placed between the target machines (the datacenter machines) and the mobile devices can perform much of the functionality of parsing monitored data and creating actionable information out of them. Related to this is the necessity for intelligently partitioning the systems management rules among the various target machines and the intermediate server. {\em Second}, we need a rich subscription language specialized for expressing events of interest for systems management, \eg, patterns in metric values, either individually or together with another metric. By doing a survey of system administrators at our institutions, we have discerned that the current simple subscriptions supported by monitoring tools are inefficient to express many common systems administration tasks. {\em Third}, we need efficient processing and communication primitives respectively for processing the event streams at the intermediate server and communicating information with the mobile devices. 

We develop an early prototype that incorporates the basic building blocks of each of the three design elements laid out above. We call this system \appname. While designing it, we find that there is a natural, albeit subtle, interplay between the two thrusts of existing work --- semantically rich subscription processing system and streaming event processing.
For evaluation, we compare \appname to the current state-of-the-art mobile monitoring solution, the highest rated application from the Google Play Store, called {\em aNag}, which is billed as a Nagios client for Android devices.
Our evaluation brings out two insights. First, \appname can deliver alerts to a mobile device in less than 10 seconds end-to-end with 231K subscriptions in the system.
Second, the aNag solution consumes high bandwidth (2 MB/minute) while monitoring just 130 target machines.

\section{Problem Background}
\label{sec:background}

Datacenter management involves continuous monitoring of system resources.
We discern that there are mainly two types of subscriptions -- (a) {\em local subscriptions} confined to a single machine, \eg, check if the CPU load on the node whose ID is 100 is greater than 80\%, and (b) {\em spatial subscriptions} which span a set of machines, \eg, check if the CPU load on each of the nodes in a subnet is greater than 80\%. Spatial subscriptions form an AND relation and are only matched when all the nodes in the subscription meet the subscription criteria.
Further, subscriptions can be based on instantaneous snapshots of metric values, or use a time window of metric values. 

Modern datacenters typically monitor six to twelve metrics per datacenter node, and typically have ten to thirty system administrators, at large academic institutions (lower end of the range) and commercial organizations with data centers for internal operations (as opposed to having it as a line of business). 
Further, it is common for system administrators to create multiple subscriptions for a metric as they might have overlapping interests \cite{rellermeyer2012system, bagchi2013lilliput}. For instance, one system administrator may be interested in the behavior of the hypervisors on a rack of machines, together with the temperature profile of the rack. A second administrator may be interested in the CPU load of the machines, along with the temperature profile.
A representative value for number of subscriptions can be arrived at by considering the following values. Each system administrator would create at most 2 subscriptions per metric, there are 10 metrics per node, and 30 system administrators giving 600 subscriptions for each physical machine. 
Also, there are VMs resident on each physical machine and for the purposes of monitoring, they are considered valid target end points. Thus, with 10 VMs per physical machine say, there would be 6,600 subscriptions, since metrics come from the 1 physical hardware/hypervisor plus 10 VMs. At ITaP, Purdue's IT organization, the number of virtual machines per host machine varies between 10 and 15. The point is that the number of subscriptions that must be supported by a middleware system tends to grow rapidly with the number of machines in a datacenter.  


\noindent{\bf A Motivating Example}

On examining the metric data from Purdue's IT organization we realized that complex subscriptions could help detect alerts in some interesting situations.
One observed issue was that the load across a set of machines behind a load balancer starts to deviate. This happens because the load balancer has some complex assignment strategies, such as, based on the number of active connections, with weights that are re-adjusted at runtime. A race condition in the assignment strategy causes the load across the machines to get disbalanced. This can be detected only though a mix of spatial monitoring (across the multiple machines which are behind the load balancer) and temporal monitoring (observing the deviation in loads grow larger with time). 


\takeout{
}


\section{Requirements for datacenter management with mobile devices}

The datacenter monitoring demands of today mean that not all monitoring can be done on the target end points (takes away resources from running the actual workload), not all monitoring can be done on the mobile devices (too much resource usage on the resource-constrained devices), and not all monitoring can be accomplished with simple instantaneous rules.

\subsection{Current Approaches}
There are a small but growing number of mobile applications that allow system administrators to manage physical target machines from the mobile devices. For example, as of March 8, 2016, there are 14 apps for systems management on the Google Play Store with ratings of 4 stars (out of 5) or higher. 

Current mobile management solutions cause high traffic between end users' mobile devices and the monitored host machines within the data center. This is fundamentally because they do not interpose any filter between the raw monitoring data generated by the target machines and the mobile devices which consume this raw data. For example, aNag, the highest rated mobile monitoring solution, which uses Nagios, consumes 1.5 MB/min for monitoring 100 target end points.
It pulls the raw data directly from the target machines \cite{carroll2010analysis}. There are some plugins with Nagios that can push the data out, but these still do not solve the problem of selectively filtering the data using a rich subscription language. Furthermore, current solutions have been directly integrated into architectures which were primarily built for desktop monitoring. For example, a plot with 5 lines in different colors may be discernible on a desktop but is not on a mobile screen.

Regarding usability, systems management tasks often require contextual information to perform the root cause analysis of failures \cite{chen2002pinpoint, bronevetsky2012automatic}. Many mobile management solutions provide only pinpointed alert information, without any context information. For example, aNag can provide 
an alert when a Nagios trigger condition is met on a remote machine, \eg, CPU utilization reaching a threshold. However, some contextual information is important, \eg, it would be important to know what the CPU utilization was on comparable machines serving the same customer's workload.

\subsection{Requirements}
To inform the design of our system monitoring application, we conducted a survey of system administrators at the central IT organization of our two institutions.
Through this, we identified the following as the primary requirements from a mobile systems monitoring solution. 
(i) The solution should enable users to receive timely alerts on a mobile device, provided they are connected to the network. If the user is not connected to the network, an asynchronous notification should be delivered when she does connect. (ii) It should be scalable to at least a few thousands of end hosts for targeting reasonable-sized commercial data centers. (iii) The solution should be parsimonious in the resource usage at the mobile devices, the most relevant resources being wireless bandwidth, computation, and resultantly, battery life. (iv) The solution should allow administrators to create on-the-fly subscriptions without having to modify individual configuration files on target machines. Such on-the-fly subscriptions are needed because system administrators realize the need for new rules as the datacenter operates and they see new faults. Making changes to the target machines is often considered a very sensitive operation because it risks disrupting the workload running on the target machines. (v) It should allow system administrators to obtain the context information for an alert, enabling them to diagnose the failure.
\section{Architecture of a mobile monitoring solution: \appname}

\begin{center}
\begin{figure}
   \includegraphics[width=\columnwidth]{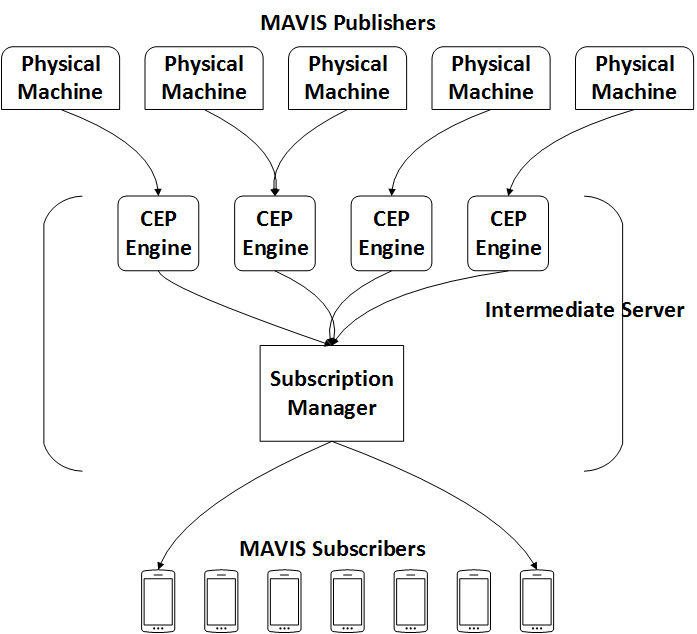}
   \caption{Flow of events in \appname during event detection}
   \label{fig:system}
\end{figure}
\end{center}

Driven by the requirements and the insights from looking at gaps with existing desktop-based monitoring solutions, we come up with an architecture for our proposed mobile-based monitoring solution, called \appname. Figure~\ref{fig:system} presents an architectural diagram of the components in \appname.
\appname comprises of the following components -- \appname publisher, \appname intermediate server which is composed of Complex Event Processing (CEP) engine(s) and a subscription manager, and the \appname subscriber, which is a mobile application. There is a \appname publisher running on each datacenter node, and several \appname publishers are associated with one \appname CEP engine (there are multiple CEP engines for the entire system). The \appname mobile application creates subscriptions and these are sent to the \appname subscription manager. Once a subscription is created on the \appname subscription manager, it is handed off to the associated \appname CEP engine(s). The process of matching collected metrics against current subscriptions either happens at the \appname publisher or the \appname CEP engine. 

\noindent{\bf \appname Domain Specific Language}

\appname offers a rich DSL which can be utilized by the \appname subscriber to create alert subscriptions. We categorize the subscriptions as {\em local} or {\em spatial} subscriptions. A second orthogonal dimension is {\em instantaneous} or {\em temporal} subscriptions. An instantaneous subscription is based on a single snapshot in time while a temporal subscription is based on metric values over a window of time. 
For temporal subscriptions, we support a variety of aggregation operators --- min, max, mean, median, standard deviation, variance, percentile, etc. Some of these operators are cheap to calculate, in terms of processing and memory, such as, min and max, while others require more resources, such as, storing more state for variance. 

\noindent{\bf Subscription Creation}
\label{sec:sub-creation}
When the mobile application creates a subscription, the subscription data is stored in the mobile device and is sent to the \appname subscription manager.
The mobile application enables users to create new subscriptions on-the-fly, and supports the dynamically changing needs of system administrators. 
The \appname subscription manager is mainly responsible for storing all the incoming subscriptions from the subscribers. Mobile applications interact with the subscription manager by subscribing for events written in the DSL. The subscription manager can be configured to create groups that contain a set of nodes within the datacenter. This becomes useful when subscribers try to create subscriptions for a large set of machines in the datacenter. The subscription manager is responsible for creating individual subscriptions for each machine in the group. 

The subscription manager can decide which CEP engines will do selective offloading of some subscriptions to the particular target node. The rationale is that all the information for matching the subscription is locally available at that node and the computational resource needed to process the subscription against the metric values is minimal. So, rather than incur the expense of communicating the stream of metric values from the publisher (the target machine) to the intermediate server, \appname decides to let the processing happen on the publisher itself. \appname can control the resource utilization on the publisher, which is a crucial setting in most systems. It does this by only offloading the easy-to-process subscriptions (such as, min and max) and bounding the total number that will be processed on any target machine.

\noindent{\bf Streaming Event Processing}
The \appname CEP engine is primarily responsible for matching subscriptions and detecting events. The event detection engine is based on IBM's InfoSphere Streams \cite{ibm-streams} stream processing engine. The use of Streams is motivated by the observation that for many applications of systems management, it is hugely inefficient to store the monitored data in a datastore and then run queries against the datastore. This is because fundamentally the queries are on data as it streams forth from the target machines and the queries are on windows of the data that are constantly sliding.

\noindent{\bf Event Detection}

In \appname, the process of detecting events and forwarding them to subscribers is split up into four parts -- metric collection, subscription matching, selective storage, and selective notifications.
Simply put, metric collection collects metrics at the target machines and can use any of several existing tools like Nagios or Ganglia. In subscription matching, the metrics are passed on to the appropriate CEP engines to then pass them through the Streams engine. With distributed CEP engines, matching may happen partially at a given engine. In selective storage, if a subscription is matched, then {\em all} the metrics for the involved machines are stored for a policy-defined amount of time (24 hours in our current setting). This is so that administrators can query and retrieve this data during a subsequent diagnosis phase following the alert. Once a subscription is matched, it normally results in an alert, \ie, a notification to the corresponding subscriber(s). \appname intermediate server sends an alert to the subscriber the first time it gets matched. If it continues to get matched, we do not send the subsequent alerts. We find the Google Cloud Messaging (GCM) system's push feature a useful service for sending the alert.

\section{Evaluation}

\subsection{Experimental Setup}

In order to conduct the evaluation of our system, we use monitoring data obtained from the host machines within ITaP, Purdue's central IT organization. ITaP monitored data is available for approximately 3,000 different machines, where each machine has between 6 and 12 metrics being monitored on them.
We use as a basis for the rules the results of our survey of system administrators. The machines and the datacenter are considered to have leading monitoring and management practices and boast of uptimes of greater than 95\% for all the clusters in 2015 \cite{purdue-community-cluster}. When virtualization is enabled, each physical machine supports 10 VMs. We obtain 4 hour continuous streams of metric data from the production machines, accumulated over 1 month in 2015.
We monitored the following 7 metrics -- User CPU, System CPU, used disk space, free memory, memory buffer space used, entropy, and ambient temperature.
 

\subsection{\appname Latency of Event Detection}
\begin{figure}
   \includegraphics[width=\columnwidth]{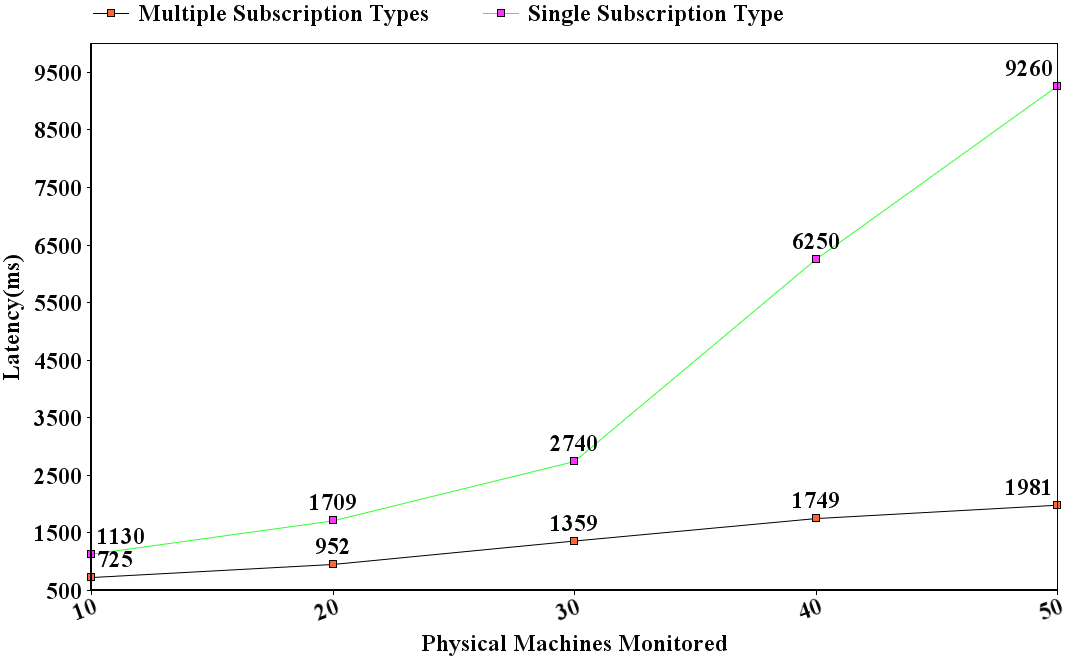}
   \caption{Average latency for receiving alerts at the \appname subscriber (the mobile device) as a function of number of target machines being monitored}
   \label{fig:latency}
\end{figure}

In this experiment we measure the end-to-end latency taken by our entire system to deliver a notification to the mobile device. We start the timer when the publisher receives the metric data on the target machine, and stop the timer when the subscriber receives the notification. In order to measure the performance of our system, we vary the number of machines being monitored and inject synthetic failures in the data. Each physical machine has 10 virtual machines. \appname monitors the metrics of the physical machines as well as the VMs. In Figure \ref{fig:latency} we show how the latency of receiving alerts at the mobile end points varies based on how we utilize the streaming event processing engine. Through the use of IBM Streams, we are able to divide the stream processing among different Streams applications. We divide the work in two ways --- (a) splitting up the total number of physical machines monitored among two Streams applications, and (b) splitting up the subscription types for {\em all} of the monitored physical machines among three Streams applications. Note that a value of 50 physical machines being monitored (X-axis of Figure \ref{fig:latency}) means there are a total of 550 end points (VMs + physical machines) being monitored.
We created approximately 4000 subscriptions per publisher. 
We injected synthetic failures in a uniform random manner in the data, such that 2\% of the overall number of subscriptions per machine would get matched.
For this experiment, publishers send the metric data to the CEP engine every 15 seconds. 

We observe the effect of a varying number of machines on the end-to-end latency (Figure \ref{fig:latency}). 
The first way of decomposing into Streams applications (by the physical machines) shows that processing all of the subscription types on a single Streams application does not scale well, since the subscription types have to be processed sequentially. Splitting up the subscription types allows each subscription type to be processed in parallel and minimizes the latency. 
In evaluating the response time of our system, scaling up to a ceiling of 550 target end points created a latency of approximately 2 seconds, which seems tolerable for most system management functions. 

\subsection{aNag Bandwidth Utilization}
Through researching existing mobile solutions for systems management, we came across aNag, the highest rated systems management application on Google Play (rating of 4.7/5). aNag is a mobile application which communicates with the Nagios monitoring software running on the target machines. 
Nagios only supports simple instantaneous subscriptions. aNag, in order to support richer subscriptions, communicates with the Nagios web interface to retrieve {\em all} the monitoring data.

 We monitored up to 130 target end points (all VMs in this case) using aNag where each target machine was monitoring between 6 and 12 metrics, and there was one subscription per metric. 
 Examining the result of the bandwidth utilization of aNag as a function of the number of target machines monitored, we find that expectedly it grows linearly, with 550 KB/s for 40 target machines and growing to 2 MB/s for 130 machines. Using \cite{bandwidth} as a reference for how bandwidth hungry popular applications are, we find that aNag exceeds the average mobile application bandwidth utilization per hour (10.7 MB/hour averaged across the 50 most popular mobile applications) after monitoring just 15 target machines, and surpasses the hungriest bandwidth usage limit (115 MB/hour for the hungriest of these 50 applications) while monitoring just 130 target machines. aNag has to retrieve all the monitoring data from the target machines, regardless of the status of the Nagios subscriptions, and filters them at the mobile application. Hence, we observe a large bandwidth utilization at the mobile device running aNag. Furthermore, due to the large number of network connections made by aNag, the mobile application utilizes an excessive number of threads and runs out of virtual memory for additional threads when monitoring more than 597 physical machines. 

\appname's utilization of a CEP engine to deliver alerts and its selective notification algorithm prevents the mobile application from using up excessive bandwidth. Notifications are only received when a {\em new failure} arises in one of the target machines.
The bandwidth utilized on sending a push notification is approximately 500 bytes, since we only send information regarding the threshold that matched the subscription, and the time at which it occurred. Administrators can also choose to query our selective storage component to analyze the failure. Querying the metrics for a VM and physical machine over the past 10 minutes, consumes 4,262 bytes. As a result, \appname subscribers consume nearly 4 KB of bandwidth when a new failure is observed.

\section{Conclusion}

The ubiquitous availability of mobile phones and widespread network connectivity have led to a demand for real-time information about the state of data centers. 
System administrators would like access to timely alerts and richer context information on the mobile devices while minimizing the data monitoring overhead on data center host machines, minimizing resource consumption on the mobile devices, and meeting the dynamically changing needs of system administrators.
We find that these requirements are incompletely met by today's monitoring solutions, which are geared toward clients running on desktop-class machines. This motivated our architecture of \appname, a framework that facilitates managing datacenter activities from mobile devices by implementing a rich domain specific language and a scalable complex event processing architecture and by integrating it with three novel components.
Our evaluation shows that the \appname can manage the activities of 3,000 physical machines within our datacenter by utilizing 6 CEP engines. A single instance of the \appname CEP engine running on one machine can handle 231K subscriptions. In contrast, the leading mobile systems management app, aNag, consumes 2 MB/s for managing just 130 machines.

{\footnotesize
\bibliographystyle{abbrv}
\bibliography{biblio}

\begin{thebibliography}{10}

\bibitem{bagchi2013lilliput}
S.~Bagchi, F.~Arshad, J.~Rellermeyer, T.~Osiecki, M.~Kistler, and A.~Gheith.
\newblock Lilliput meets brobdingnagian: Data center systems management through
  mobile devices.
\newblock In {\em The Third International Workshop on Dependability of Clouds,
  Data Centers and Virtual Machine Technology (DCDV)}, pages 1--6. IEEE, 2013.

\bibitem{bronevetsky2012automatic}
G.~Bronevetsky, I.~Laguna, B.~R. de~Supinski, and S.~Bagchi.
\newblock Automatic fault characterization via abnormality-enhanced
  classification.
\newblock In {\em 42nd Annual IEEE/IFIP International Conference on Dependable
  Systems and Networks (DSN)}, pages 1--12. IEEE, 2012.

\bibitem{carroll2010analysis}
A.~Carroll and G.~Heiser.
\newblock An analysis of power consumption in a smartphone.
\newblock In {\em USENIX annual technical conference}, pages 271--285, 2010.

\bibitem{chen2002pinpoint}
M.~Y. Chen, E.~Kiciman, E.~Fratkin, A.~Fox, and E.~Brewer.
\newblock Pinpoint: Problem determination in large, dynamic internet services.
\newblock In {\em Dependable Systems and Networks, 2002. DSN 2002. Proceedings.
  International Conference on}, pages 595--604. IEEE, 2002.

\bibitem{meraki-systemsmanager}
{Cisco Systems}.
\newblock {Meraki Systems Manager}.
\newblock \url{https://play.google.com/store/apps/details?id=com.meraki.sm}.

\bibitem{aNag}
{Damien Degois}.
\newblock {aNag}.
\newblock
  \url{https://play.google.com/store/apps/details?id=info.degois.damien.android.aNag}.

\bibitem{nagios}
{Ethan Galstad}.
\newblock {Nagios - The Industry Standard in IT Infrastructure Monitoring}.
\newblock \url{http://www.nagios.org/}.

\bibitem{ganglia}
{Ganglia Community}.
\newblock {Ganglia Monitoring System}.
\newblock \url{http://ganglia.sourceforge.net/}.

\bibitem{ibm-streams}
IBM.
\newblock {Big Data: Stream Computing}.
\newblock \url{http://www.ibm.com/developerworks/bigdata/stream.html}.

\bibitem{purdue-community-cluster}
{Information Technology at Purdue (ITaP)}.
\newblock {Community Clusters}.
\newblock \url{https://www.rcac.purdue.edu/services/communityclusters/}.

\bibitem{bandwidth}
{Kevin Tofel}.
\newblock {Data Hungry Mobile Apps Eating into Bandwidth Use}.
\newblock
  \url{http://gigaom.com/2011/05/19/data-hungry-mobile-apps-eating-into-bandwidth-use/}.

\bibitem{rellermeyer2012system}
J.~S. Rellermeyer, T.~H. Osiecki, E.~A. Holloway, P.~J. Bohrer, and M.~Kistler.
\newblock System management with ibm mobile systems remote: A question of power
  and scale.
\newblock In {\em Mobile Data Management (MDM), 2012 IEEE 13th International
  Conference on}, pages 294--299. IEEE, 2012.

\end{thebibliography}
}
\end{document}